\input stromlo

\title Optical Synthetic Spectra of Elliptical Galaxies

\shorttitle Optical Synthetic Spectra

\author Roger A. Bell and Mark L. Houdashelt

\shortauthor Bell \& Houdashelt

\affil Department of Astronomy, University of Maryland

\maintext

We present the first results from our population synthesis models of elliptical
galaxies.  Here, we concentrate upon the optical region of the integrated
spectrum (3000--7000 \AA).
A companion paper in this volume describes the models more fully and presents a
discussion of the near-infrared (0.7--3 $\mu$m).

Figure 1 compares our 16 Gyr, [Fe/H]=0.0 model spectrum to
Kennicutt's (1992) spectrum of the E1/S0 galaxy NGC 4472.  Since
Kennicutt warns that his spectrophotometry is only accurate to $\sim$10\%
over the breadth of the wavelength region shown, the
similarities between the two spectra are encouraging.

\figureps[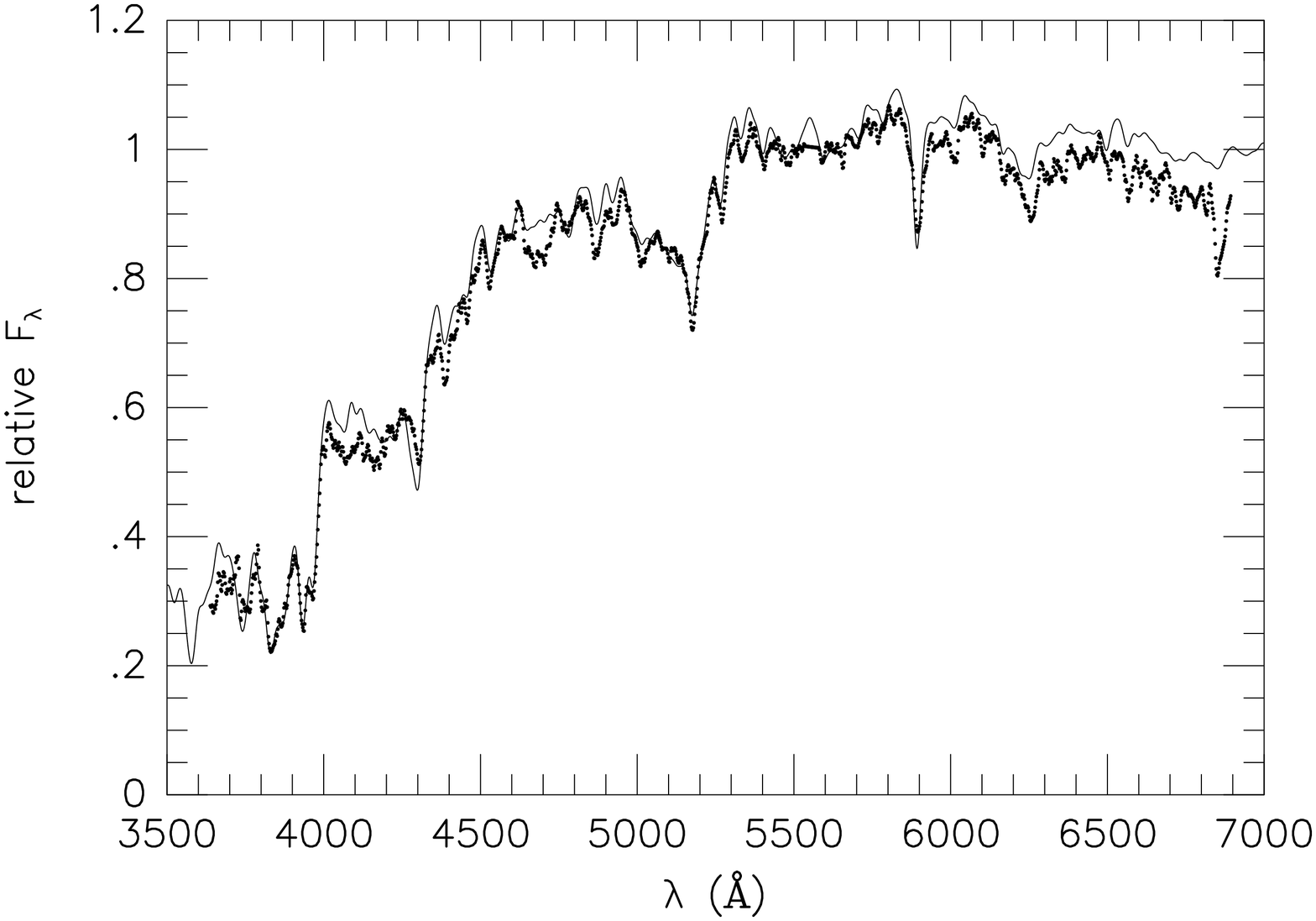,.7\hsize] 1. A comparison of the spectrum of
NGC 4472 (Kennicutt 1992; solid points) to our synthetic spectrum for a
16 Gyr, [Fe/H]=0.0 stellar population (solid line).  The spectra are
normalized at 5500 \AA, and the synthetic
spectrum has been convolved and rebinned to the resolution of Kennicutt's data.

Figure 2 shows how the optical colors
measured from our synthetic spectra vary with the age and metallicity of the
stellar population.  Worthey (1994; hereafter W94) model colors
and some photometry of E/S0 galaxies are also shown.
Our models are more consistent with the color-color relations of E galaxies
being caused by variations in metallicity rather than age changes.
Our 16 Gyr models infer
that the reddest ellipticals have [Fe/H]$\sim$+0.25.
W94 predicts a lower metallicity because his stellar continua
are $\sim$0.06 mag too red in $B-V$.

\figureps[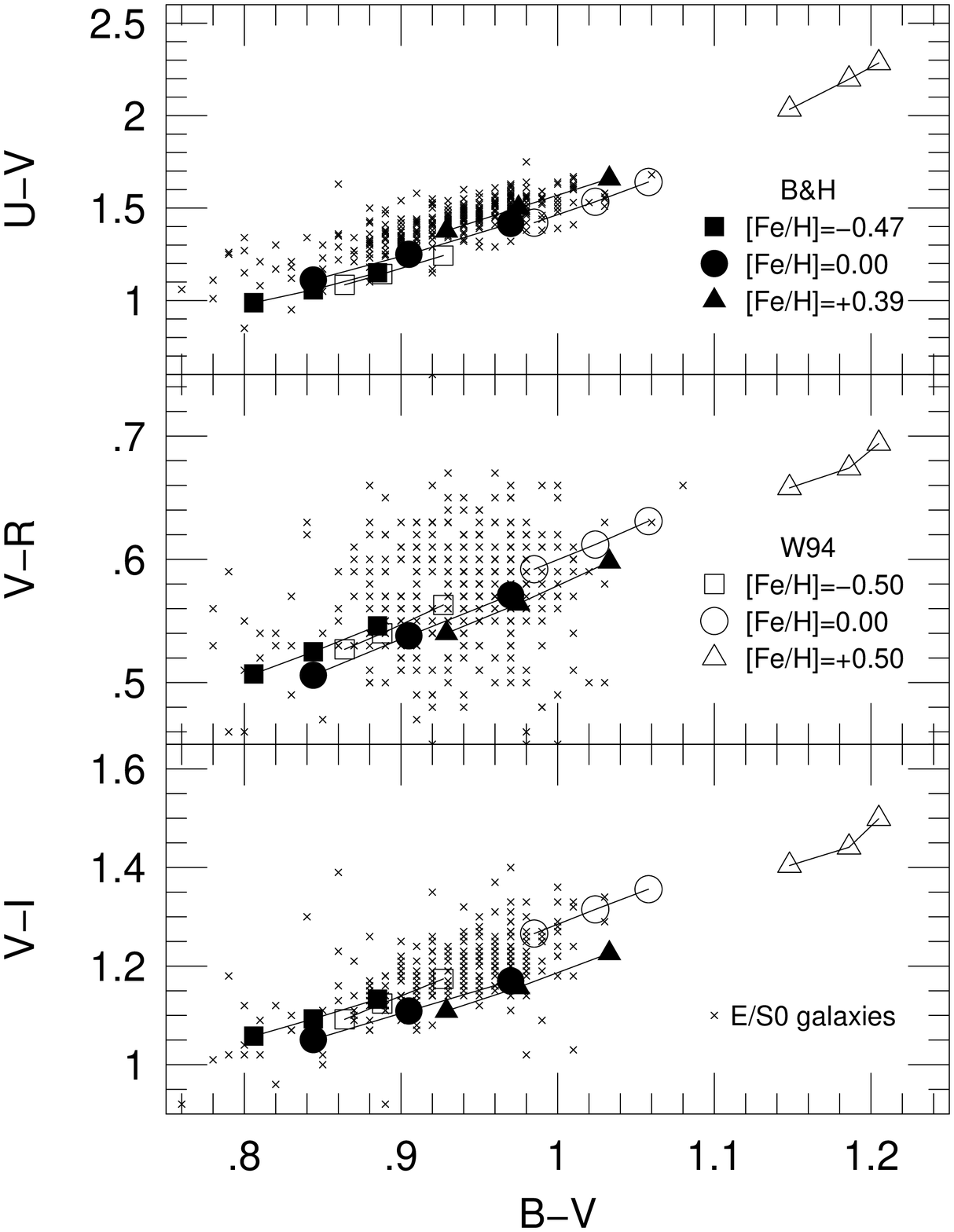,.59\hsize] 2. The optical colors measured from
our synthetic spectra are compared to the models of W94
and photometry of E/S0 galaxies (Prugniel \& Simien 1996).
From blue to red at each metallicity, our 6, 10 and 16 Gyr models and the
8, 12 and 17 Gyr models of W94 are compared.

Our model Lick indices (not shown) follow the galaxy
trends about as well as the W94 models.
However, the age/metallicity associated with
a specific color or index in the two sets of models usually differ.
While we do not have the
problem that W94 encountered in matching the TiO$_1$, TiO$_2$ relation
of elliptical galaxies, our models do not exhibit the same
CN$_1$, CN$_2$ and Mg$_2$, Mg {\it b} trends as the W94 models, which
closely follow the galaxy measurements.
Since CN$_1$ and CN$_2$ are defined by the same CN bandpass, the difference
must lie in the continuum bands used to measure the indices (perhaps in the
strength of H$\delta$).  We suspect that some factor other than Mg abundance
is causing the Mg$_2$, Mg {\it b} discrepancy
(see Tripicco \& Bell 1995).  Surprisingly, neither set of models is
able to simultaneously match the Mg$_2$ and Mg$_1$ indices of E
galaxies.

This work was supported by NASA Grant NAG53028 and NSF Grant AST93-14931.

\references

Kennicutt, R.C., Jr. 1992, ApJS, 79, 255.

Prugniel, P. \& Simien, F. 1996, A\&A, 309, 749.

Tripicco, M.J. \& Bell, R.A. 1995, AJ, 110, 3035.

Worthey, G. 1994, ApJS, 95, 107. (W94)

\bye